\documentclass[%
 reprint,
unsortedaddress,
 amsmath,amssymb,
 prl,
]{revtex4-1}

\usepackage{dsfont}
\usepackage{graphicx}
\usepackage{hyperref}
\usepackage{color}

\usepackage{mathrsfs}
\usepackage{braket}

\usepackage{overpic}
\usepackage{tikz}
\usetikzlibrary{decorations.pathreplacing}

\newcommand{\im}{\mathrm{i}}

\definecolor{mygreen}{rgb}{0,0.5,0}
\definecolor{myblue}{rgb}{0,0,0.75}
\definecolor{mymagenta}{cmyk}{0,1,0,0.12}

\usepackage{umoline}

\begin{document}
\title{Irreversible dynamics in quantum many-body systems}

\author{Markus Schmitt}
\email{markus.schmitt@theorie.physik.uni-goettingen.de}
\author{Stefan Kehrein}
\affiliation{%
 Institute for Theoretical Physics, 
	Georg-August-Universit\"at G\"ottingen, 
	Friedrich-Hund-Platz 1 - 37077 G\"ottingen, Germany
}
\date{\today}

\begin{abstract}
Irreversibility, despite being a necessary condition for thermalization, still lacks a sound understanding in the context of isolated quantum many-body systems. In this work we approach this question by studying the behavior of generic many-body systems under imperfect effective time reversal, where the imperfection is introduced as a perturbation of the many-body state at the point of time reversal. Based on numerical simulations of the full quantum dynamics we demonstrate that observable echos occurring in this setting decay exponentially with a rate that is independent of the perturbation; hence, the sensitivity to perturbations is intrinsic to the system meaning that the dynamics is effectively irreversible.
\end{abstract}

\maketitle

\paragraph{Introduction.}
The recent development of experimental techniques to realize and precisely manipulate closed
quantum systems with many degrees of freedom 
\cite{Greiner2002, Kinoshita2006,Choi2016,Jurcevic2017,Zhang2017,Bernien2017,Guardado-Sanchez2018} 
motivated a lot of theoretical activity aimed
at understanding the dynamics of quantum many-body systems far from equilibrium.
A fundamental question that arose in this context is how and in what sense closed 
quantum many-body systems thermalize when initially prepared far from
thermal equilibrium. This has been investigated with great efforts in recent years 
\cite{Eisert2015,dAlessio2016}.
Closely related is the question of irreversibility, which, however, did to date not receive as much attention; in the context of quantum many-body systems different notions of irreversibility are under discussion \cite{Levstein1998,Usaj1998,Zangara2015,Schmitt2016,Swingle2018}.

For classical systems the origin of irreversibility despite microscopically reversible dynamics 
 was already discussed by Boltzmann and Loschmidt
\cite{Boltzmann1872,Loschmidt1876,Boltzmann1877} and was 
essentially understood in a modern sense by Thompson \cite{Thompson1874}.
Classical systems typically exhibit chaotic dynamics if composed of many degrees of freedom.
Hence, any practical efforts to revert the dynamics, e.g. by inverting the momenta, 
are ultimately futile due to the exponential sensitivity of the dynamics to small imperfections.
In particular, the dominant rate with which initially nearby trajectories diverge, called Lyapunov exponent,
is independent of the perturbation strength. Therefore, any improvement of the accuracy in the time
reversal protocol can only affect the prefactor of the exponential law.

This practical understanding of irreversibility in classical systems 
led Peres \cite{Peres1984} to introduce the Loschmidt echo
\begin{align}
	\mathcal L(t)=|\langle\psi_0|e^{\im (\hat H+\epsilon\hat V)t}e^{-\im\hat Ht}|\psi_0\rangle|^2
\end{align}
as measure for irreversibility in quantum systems. The Loschmidt echo is the overlap
of a wave function evolved forward in time with Hamiltonian $\hat H$ and subsequently backwards
with a slightly perturbed Hamiltonian $\hat H+\epsilon\hat V$, thereby quantifying the resemblance of the
time evolved state with the initial state. This quantity proved very useful in the analysis of the
dynamics of quantum systems with few degrees of freedom \cite{Gorin2006,Jacquod2009}.

\begin{figure}[t]
\includegraphics{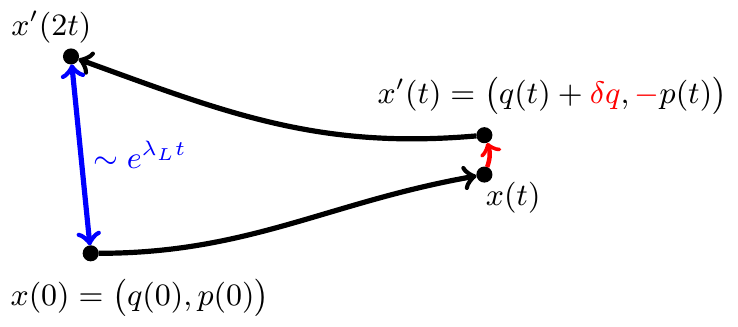}
\caption{Classical analog of the time reversal protocol under consideration.}
\label{fig:class_traj}
\end{figure}

In generic quantum many-body systems, however, overlaps like the Loschmidt echo have only
limited significance for the resemblance of states in physical terms. According to the
eigenstate thermalization hypothesis (ETH) \cite{Deutsch1991,Srednicki1994,Srednicki1996,Rigol2008} 
energy expectation values of few-body observables 
$\mathcal O_E=\langle E|\hat O|E\rangle$ are smooth functions of the eigenstate energy $E$. 
More precisely, the assumption is that the difference of the expectation value in neighboring eigenstates, 
$|\mathcal O_{E_n}-\mathcal O_{E_{n+1}}|$, is exponentially suppressed with increasing
system size, which is strongly supported by numerical evidence from different studies 
\cite{Steinigeweg2013,Beugeling2014,Kim2014,Mondaini2016}.
Since all experimentally measurable quantities are related to the above-mentioned class of observables 
this means that energetically
close-by eigenstates of a generic Hamiltonian $\hat H$, although orthogonal, 
are by all practical means indistinguishable in experiment. 
The same holds for
integrable systems if the further integrals of motion are taken into account in addition to the energy 
\cite{Cassidy2011}. Therefore, any definition of irreversibility in many-body systems should be
based on observables, which are accessible in experiment \citep{Levstein1998,Usaj1998,Fine2014,Zangara2015,Elsayed2015,Schmitt2016,Fine2017}. 

Connected to the question of irreversibility 
so called out-of-time-order correlators (OTOCs) of the form $\langle \hat A^\dagger(0)\hat B^\dagger(t)\hat A(0)\hat B(t)\rangle_\beta$ were recently suggested to probe scrambling, i.e. the
complete delocalization of initially local information, and exponential sensitivity of the dynamics to small
perturbations \cite{Shenker2014,Kitaev2014}. Based on this a black hole
theory and a holographic model of Majorana fermions were identified as maximally chaotic systems \cite{Maldacena2016,Kitaev2015,Maldacena2016a}. Moreover, OTOCs can directly be related to an information-theoretic 
measure for the delocalization of initially local information \cite{Hosur2016}.
These ideas were seized theoretically in a number of subsequent works to investigate signatures of 
chaos and scrambling in the dynamics of local lattice models 
\cite{Hosur2016,Bohrdt2016,Roberts2016,Iyoda2017,Swingle2016,Shen2016,Huang2016,Patel2017,vonKeyserlingk2017,Rakovszky2017,He2017,Lin2018,Bordia2018,Khemani2018,Nahum2018} and first experiments were conducted \cite{Garttner2017,Li2017,Meier2017,Wei2018}.

\paragraph{Scope of this work.}
In this work we propose a probe of irreversible dynamics based on observable echos under imperfect
effective time reversal. A first investigation in an integrable model was done in Ref. \cite{Schmitt2016}.
There, it was found that the echos in an integrable spin chain decay not faster than algebraically
indicating that the dynamics is well reversible. Under a certain protocol even ever-persisting echos were
found.
In the present work we demonstrate that in generic non-integrable systems 
the observable echos under imperfect time reversal decay
exponentially as would be expected when the dynamics is chaotic and -- importantly -- that the decay is 
primarily governed by the intrinsic properties of the system. This finding contrasts the 
aforementioned algebraic decay found in an integrable spin chain.
It is, moreover, in contrast to protocols involving a perturbation of the Hamiltonian, where the decay timescale naturally depends on the perturbations strength \cite{Zangara2015,SchmittThesis2018}.
In the considered protocol the imperfection is introduced as a perturbation of the
many-body state at the point of time reversal. Hence, the proposed probe of irreversibility 
directly corresponds with 
the understanding of classical irreversibility as a consequence of the
butterfly effect, but it is applicable to generic quantum many-body systems far from any semi-classical
limit.

\paragraph{Time reversal protocol.}
Irreversibility in classical systems is understood to be a consequence of chaotic dynamics,
i.e. the fact that trajectories diverge exponentially if the coordinates are slightly changed initially.
This leads to the fact that the final coordinates deviate exponentially from the initial coordinates
if an imperfect time reversal protocol as sketched in Fig.\ \ref{fig:class_traj} is applied.

An analogous situation in quantum systems is the perturbation of the quantum state at the point of 
time reversal, i.e., applying a unitary perturbation operator $\hat P_\epsilon$ to the time-evolved state
$|\psi(t)\rangle=e^{-\im\hat Ht}|\psi_0\rangle$,
\begin{align}
	|\psi(t)\rangle\to|\psi'(t)\rangle=\hat P_\epsilon|\psi(t)\rangle\ ,
\end{align}
where $\epsilon$ is a parameter for the magnitude of the perturbation.
For many-body systems it is crucial to regard physical observables as measure for
the smallness of the perturbation and not the overlap of the states. Due to the unitarity of the
time evolution there will always be a part of the dynamics that is perfectly reverted if the
states before and after applying the perturbation have a non-vanishing overlap. 
A natural operation $\hat P_\epsilon$ 
that leaves observables almost unchanged while making the state orthogonal is time evolution
with a local extensive Hamiltonian $\hat H_p$ for short time $\delta t$. In a
system with $N$ degrees of freedom the return probability generally takes the form $|\langle\psi_0|e^{-\im\hat Ht}|\psi_0\rangle|^2=e^{-Nr(t)}$ with
an intensive rate function $r(t)$, i.e., the overlap vanishes at arbitrarily short times
in the thermodynamic limit $N\to\infty$. Observables, instead, change smoothly under time evolution
with a physical Hamiltonian.

In the following we study the dynamics of a many-body system when a time reversal protocol
motivated by these considerations is applied. The system is prepared in an initial state 
$|\psi_0\rangle$ that exhibits some significant features distinguishing it from an equilibrium state of the Hamiltonian $\hat H$, like, e.g., a strong magnetic order in a disordered phase. 
This state is time-evolved for a waiting time $\tau$, yielding $|\psi(\tau)\rangle=e^{-\im\hat H\tau}|\psi_0\rangle$.
At this point a perturbation operator $\hat P_{\delta t}=e^{-\im\hat H_p\delta t}$ given by some other
Hamiltonian $\hat H_p$ is applied for a short time $\delta t$, resulting in 
$|\psi'(\tau)\rangle=\hat P_{\delta t}|\psi(\tau)\rangle$. 
Subsequently, $|\psi'(\tau)\rangle$ is evolved backwards
in time until the echo time $t^*\approx 2\tau$, where the resemblance of the time evolved state
to the initial state is largest in terms of the observables under consideration, i.e., these observables
show an extremum, which we call an echo peak. The existence of these echo peaks can be inferred
by considering the case of $\delta t=0$, where a perfect revival is produced independent of the
waiting time $\tau$, and assuming a smooth behavior of the dynamics as $\delta t$ is increased.
We propose to declare a system irreversible if the decay of echos as a function of the waiting time $\tau$
is exponential or faster than exponential and if the decay rate is an intrinsic property of the system, i.e.,
unaffected by reducing the perturbation strength. This definition means that substantial improvement
of the reconstruction of the initial state by manipulating with enhanced precision is practically impossible.

Note that by identifying $\hat A(0)\equiv \hat O$ and
$\hat B(\tau)\equiv e^{\im\hat H\tau}\hat P_{\delta t}e^{-\im\hat H\tau}$ 
this protocol effectively results in the measurement of an OTOC as introduced above 
if the initial state $|\psi_0\rangle$ 
is an eigenstate of the observable under consideration, 
$\hat O|\psi_0\rangle=O|\psi_0\rangle$; see also Ref. \cite{Garttner2017}.
A key difference is, however, the fact that the echo protocol takes into account the expectation value
in the pure initial state far from equilibrium, whereas the OTOC is originally defined with respect to a thermal
density matrix \cite{Shenker2014,Kitaev2014}.

\begin{figure}[t]
\includegraphics{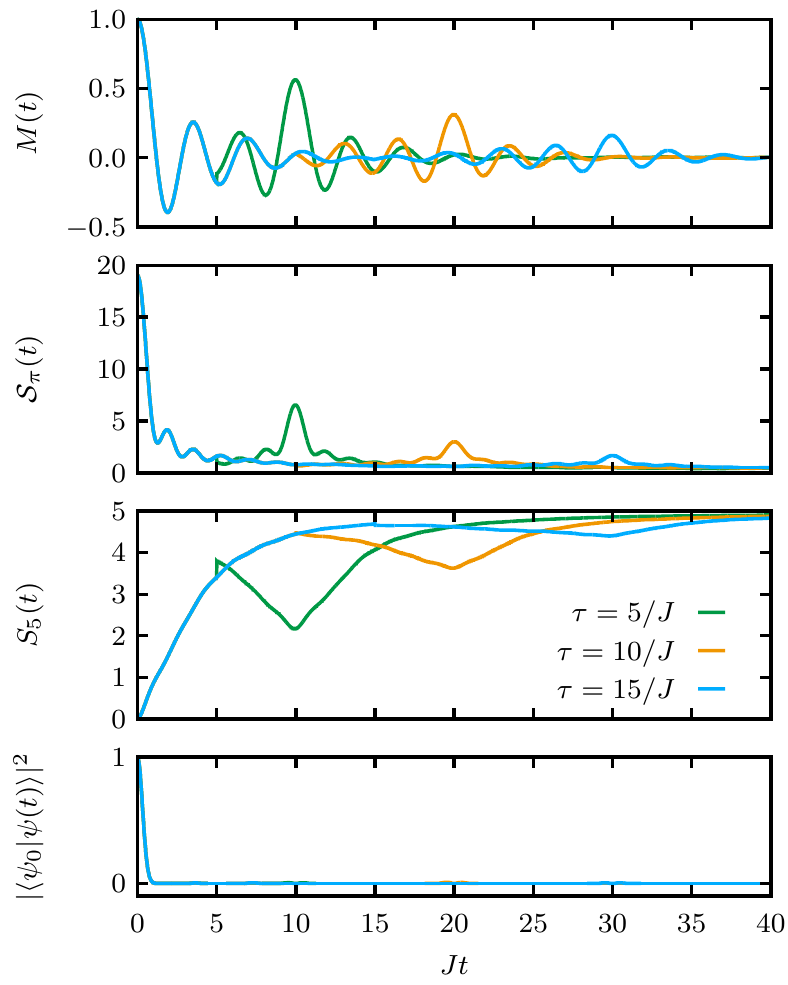}
\caption{Time evolution of staggered magnetization $M(t)$, spin structure factor $\mathcal S_\pi(t)$, entanglement entropy $S_5(t)$, and return probability $|\langle\psi_0|\psi(t)\rangle|^2$ under the
imperfect effective time reversal protocol for different forward times $\tau$ obtained with the
Hamiltonian $H_\text{loc}$. The observables and the entanglement entropy show clear
echos at $t=2\tau$ that decay as $\tau$ is increased, whereas the return probability does not 
show any signal. The perturbation strength is $\delta t/J=0.05$.}
\label{fig:full_echo}
\end{figure}
\begin{figure*}[t]
\includegraphics{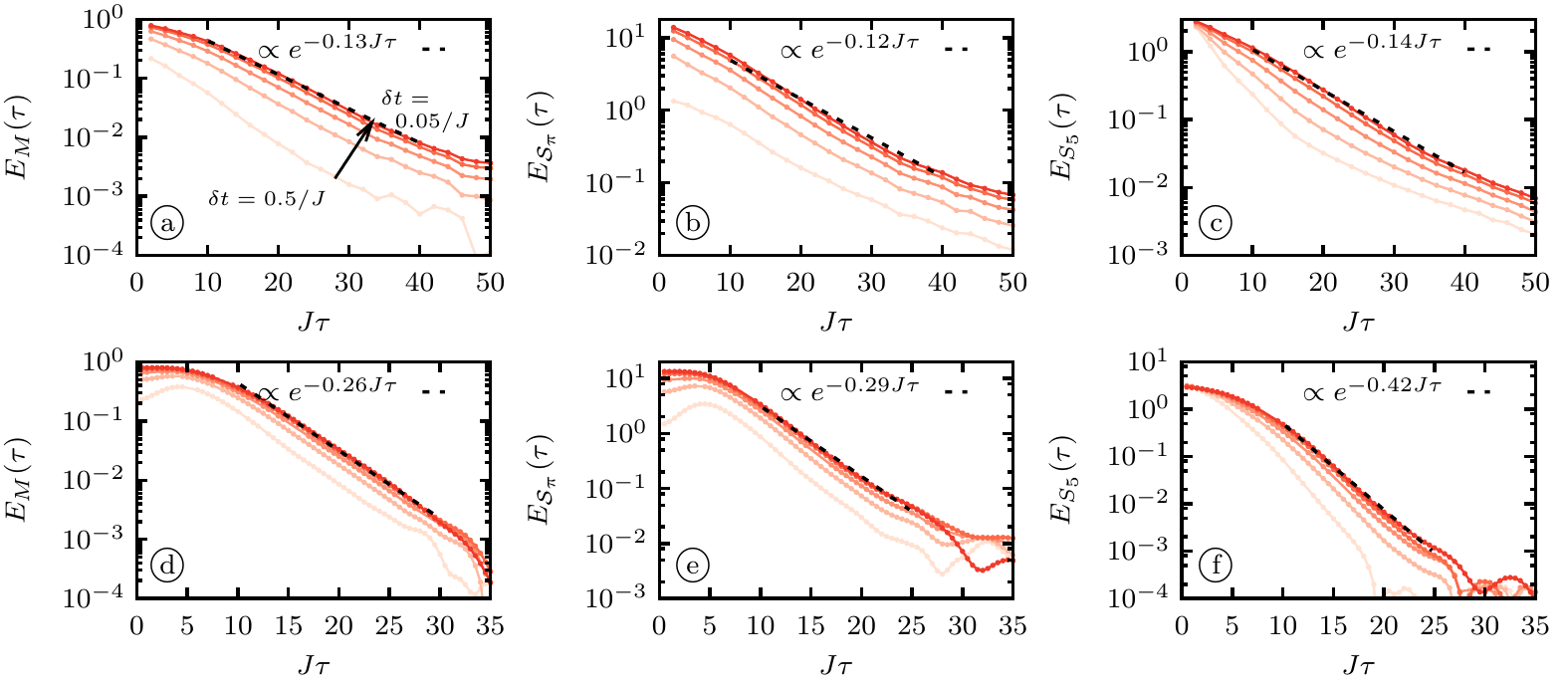}
\caption{Decay of the echo peak heights of staggered magnetization $M$, spin structure factor 
$\mathcal S_\pi$, and entanglement entropy of five consecutive spins $S_5$ 
after imperfect effective time reversal for both the local Hamiltonian $\hat H_\text{loc}$ with $N=24$ (a)-(c)
and the fully connected Hamiltonian $\hat H_\text{fc}$ with $N=22$ (d)-(f). The perturbation Hamiltonian is the same realization
of $\hat H_p$ in all cases, whereas the plotted perturbation strengths are 
$J\delta t=0.5,0.35,0.25,0.15,0.05$. The dashed lines indicate exponential fits to the results for $\delta t=0.05/J$.}
\label{fig:echo_decay}
\end{figure*}
\paragraph{Model Hamiltonians.}
As minimal examples of generic quantum many-body systems we study spin-1/2
systems defined by the Hamiltonian
\begin{align}
	\hat H=\sum_{i\neq j}J_{ij}\big(\hat \sigma_i^x\hat \sigma_{j}^x+\hat \sigma_i^y\hat \sigma_{j}^y\big)\ ,
\end{align}
where the $\hat \sigma_i^\alpha$, $\alpha=x,y$, denote the Pauli spin operators acting on 
lattice sites $i=1,\ldots,N$.
In this work we focus on two versions of this Hamiltonian, namely a Hamiltonian 
$\hat H_\text{loc}$ with local couplings 
\begin{align}
	J_{ij}^{\text{loc}}=J\left\{\begin{array}{ll}
	2^{1-|i-j|}&\text{, for }0<|i-j|\leq2\\
	0&\text{, else}
	\end{array}\right.
\end{align}
and a fully connected random Hamiltonian $\hat H_\text{fc}$ with 
$J_{ij}^{\text{fc}}=J_{ji}^{\text{fc}}=JR_{ij}/N$, where $R_{ij}$ is 
drawn from the standard normal distribution. We found that altering the interaction range of the local
Hamiltonian and introducing an (anisotropic) Heisenberg-type coupling left the results qualitatively
unchanged. However, the restriction of the couplings to shorter distances or 
strongly anisotropic couplings, respectively,
introduce large oscillations to the dynamics, which complicates the identification of echo peaks.
As initial state we choose the N\'eel state
$\ket{\psi_0\rangle=|\uparrow\downarrow\uparrow\ldots}$. For this state the staggered magnetization
$M=\frac{1}{N}\sum_n(-1)^n\langle\hat\sigma_n^z\rangle$ and the spin structure factor 
$\mathcal S_\pi=\frac{1}{N}\sum_{i,j}e^{\im (i-j)\pi/N}\langle\vec \sigma_i\cdot\vec \sigma_j\rangle$ 
constitute suited observables for the echo protocol described above. Moreover, we will investigate
the dynamics of the entanglement entropy of bipartitions into subsystems $A$ and $B$ defined by
$S_A=-\text{tr}\big(\hat\rho_A\log_2\hat\rho_A\big)$,
where $\hat \rho_A=\text{tr}_B\big(|\psi(t)\rangle\langle\psi(t)|\big)$ is the reduced density matrix of the subsystem $A$.

The Hamiltonian that defines the perturbation is chosen to be a local random Hamiltonian
\begin{align}
	\hat H_p=\sum_{i=1}^{N-1}J_{i}\big(\hat \sigma_i^x\hat \sigma_{i+1}^x+\hat \sigma_i^y\hat \sigma_{i+1}^y\big)
	\label{eq:pertham}
\end{align}
with real couplings $J_{i}$ drawn from the standard normal distribution.
However, the characteristics of the echo dynamics do not depend on this particular choice of $\hat H_p$.

\paragraph{Numerical realization and finite size effects.}
In the following we will resort to exact diagonalization and Lanczos propagation \cite{Park1986} in
order to compute the time evolution. This
limits the accessible systems to sizes far from the thermodynamic limit; due to the numerical
expense the maximal system size we consider is $N=24$.
For any finite system, however, the echos produced under the envisaged imperfect effective time reversal
will generally not decay to zero for long waiting times $\tau$. 
Introducing the eigenbasis of the Hamiltonian, $(\hat H-E_{\alpha})|\alpha\rangle=0$, the time evolution
of observables under the time reversal protocol is
\begin{align}
	\langle\hat O\rangle_{t_1,t_2}&=
	\langle\psi_0|e^{\im\hat Ht_1}\hat P_{\delta t}^\dagger e^{-\im\hat Ht_2}\hat O
	e^{\im\hat Ht_2}\hat P_{\delta t}e^{-\im\hat Ht_1}|\psi_0\rangle
	\nonumber\\&
	=
	\sum_{\alpha,\alpha',\beta,\beta'}
	\langle\psi_0|\alpha\rangle\big(P_{\delta t}^\dagger\big)_{\alpha,\alpha'} O_{\alpha'\beta}
	\big(P_{\delta t}\big)_{\beta\beta'}\langle\beta'|\psi_0\rangle
	\nonumber\\&\quad\quad
	\times e^{\im(E_\alpha-E_{\beta'}) t_1+\im(E_\beta-E_{\alpha'})t_2}\ ,
\end{align}
where $X_{\alpha\beta}=\langle\alpha|\hat X|\beta\rangle$ with 
$\hat X=\hat O,\hat P_{\delta t}, \hat P_{\delta t}^\dagger$ denotes the matrix elements of the respective operators. Clearly, for any $t_1,t_2$ terms with $\alpha=\beta'$ and $\beta=\alpha'$ are time-independent.
These terms yield the stationary value that is reached at long times $t_1\neq t_2$.
At $t_1=t_2$ there is, however, an additional time-independent contribution of the terms with
$\alpha=\alpha'$ and $\beta=\beta'$, where the diagonal elements of the perturbation operator
$\big(P_{\delta t}\big)_{\alpha\alpha}=\langle\alpha|e^{-\im\hat H_p\delta t}|\alpha\rangle$ appear.
Most prominent among these contributions at small $N$ is the identity that gives rise to the non-vanishing
overlap $\langle\psi(\tau)|\hat P_{\delta t}|\psi(\tau)\rangle$ in the finite system.
As discussed above the modulus of overlaps of the form 
$\langle\alpha|e^{-\im\hat H_p\delta t}|\alpha\rangle$
vanishes at arbitrarily short times in the thermodynamic
limit. Hence, this non-decaying contribution to echos at $t_1=t_2$ vanishes for $N\to\infty$.
In the supplemental material \cite{supplement} we demonstrate that the anticipated dependence of the persistent echos on system size can indeed be observed in our data.

In the finite systems we analyze the decay of the echo peaks towards these stationary values,
which is the universal behavior that survives in the thermodynamic limit.
In practice we discard finite size contributions by including a projection onto the subspace orthogonal to the unperturbed state with the perturbation we apply.
When analyzing the decay laws we additionally subtract the 
remaining stationary value from the echo peak heights;
see supplemental material \cite{supplement} for further explanation of the analysis of the numerical data.

Fig. \ref{fig:full_echo} displays an exemplary time evolution with $\hat H_\text{loc}$ 
of staggered magnetization $M(t)$,
spin structure factor $\mathcal S_\pi(t)$, entanglement entropy $S_n(t)$ of $n=5$ spins
at one end of the spin chain, and overlap with the initial state $|\langle\psi_0|\psi(t)\rangle|^2$ for a system 
of $N=20$ spins, where the perturbation with $J\delta t=0.05$ is applied at different waiting times $\tau$. 
Here, the time-dependent state is
\begin{align}
	\ket{\psi(t)}=
	\begin{cases}
		e^{-\im\hat Ht}\ket{\psi_0}\ ,\quad t<\tau\\
		e^{\im\hat H(t-\tau)}\hat P_{\delta t}e^{-\im\hat H\tau}\ket{\psi_0}\ ,\quad t>\tau\\
	\end{cases}\ ,
\end{align}
from which $M(t)$, $\mathcal S_\pi(t)$, and $S_n(t)$ are obtained.
The perturbation causes only a minimal shift of the observables although the perturbed state
is orthogonal to the state before the perturbation and the dynamics exhibits pronounced echo peaks 
at $t_e\approx2\tau$. 
Note that in
contrast to the results for imperfect time reversal with a perturbed Hamiltonian \cite{Schmitt2016}
the echo time under the present time reversal protocol is always very close to $2\tau$ and does not 
exhibit any systematic shift away from that.

\paragraph{Echo peak decay.}
As is evident from the exemplary time evolution in Fig.\ \ref{fig:full_echo} the resemblance of the
time-evolved state to the initial state in terms of the observables decreases as the waiting time
is increased. In order to extract laws of decay we introduce the echo peak height of an observable 
$\hat O$,
\begin{align}
	E_O(\tau)
	=\max_{t'>\tau}|\langle\hat O\rangle_{t',\tau}-O_\infty|\ ,
\end{align}
where the maximum occurs at the echo time $t_e\approx2\tau$ and $O_\infty$ is the stationary value
reached after long times.

Fig. \ref{fig:echo_decay} displays the decay of the echo peak heights for the observables and the 
entanglement entropy for both the local Hamiltonian $\hat H_\text{loc}$ and the fully connected random 
Hamiltonian $\hat H_\text{fc}$. The decay for a single realization of the perturbation Hamiltonian $\hat H_p$ 
is shown for different perturbation strengths $\delta t$. In all cases the echo peak heights exhibit a
marked exponential decay at long waiting times $\tau$. The decay rate 
varies only weakly as $\delta t$ is changed. In particular, the curves converge as $\delta t\to0$.
In each plot an exponential fit to the data with the smallest perturbation strength ($\delta t=0.05$) is
included. For the local Hamiltonian the fitted decay rates for both observables and the entanglement
entropy are almost identical. In the fully connected system the observable echos decay with similar rates,
whereas the echos in the entanglement entropy decay slightly faster. 
While the relation of the decay rates to microscopic properties of the systems is as of yet unclear, 
we find that they
do not coincide with the decay rates occurring after a simple quench.
The decay of the entanglement
entropy shows that although recoverable at short times the information about the genuinely quantal 
structure of the initial state is lost in the same fashion as the information about observables.

For different realizations of the random perturbation Hamiltonian $\hat H_p$ and fixed $\delta t$ 
we observed variations of the decay rate of about 15\%. We attribute these variations to the small
system size and expect them to vanish in the thermodynamic limit.


\paragraph{Discussion.}
In this work results from numerical simulations of the full quantum dynamics are reported. 
Our results show that generic quantum many-body systems exhibit exponential decay of 
observable echo peaks under imperfect effective time reversal. This is in contrast to algebraically 
decaying echos found in an integrable system \cite{Schmitt2016}. 
Importantly, the decay rate in the non-integrable
quantum many-body models studied here was found to be largely independent of the perturbation 
strength. This implies that any practical effort to improve the accuracy in a time reversal experiment 
is in the end futile, just like in irreversible classical systems.  

The presented results give rise to further questions, which are beyond the scope of this work and 
are therefore left for future research.
It was found that the decay rate of the echos is an intrinsic property of the Hamiltonian that
determines the time evolution. However, it is at this point not clear how said rate is related to the
microscopic details of the system. 
Moreover, possible relations to other definitions of quantum chaos and irreversibility,
e.g. the one based on OTOCs \cite{Shenker2014,Kitaev2014}, should be investigated.

Regarding experimental relevance, understanding the echo decay in magic echo setups \cite{Schneider1969,Rhim1971,Hafner1996,Morgan2012} is a long standing problem directly related to the fundamental question which we are addressing in this work, although the many-body Hamiltonians relevant for magic echo problems are beyond the scope of this work \cite{Pastawski2000}. For a highly controlled experimental setup we propose quantum simulators like those based on trapped ions \cite{Monz2011,Blatt2012,Zhang2017,Jurcevic2017} or Rydberg atoms \cite{Bernien2017,Guardado-Sanchez2018}, which effectively implement the dynamics of spin models and where one would also be able to investigate the finite size behavior. 
In fact, a protocol very similar to our echo prescription has recently been realized with trapped ions to investigate the dynamics of OTOCs \cite{Garttner2017}.

\begin{acknowledgments}
This work was supported through SFB 1073 (project B03) of the
Deutsche Forschungsgemeinschaft (DFG).
M.S.\ acknowledges support by the Studienstiftung des Deutschen Volkes.
For the numerical computations the Armadillo library \cite{armadillo} was used.
\end{acknowledgments}

\end{document}


\title{Supplemental material to\\\emph{Irreversible dynamics in quantum many-body systems}}

\author{Markus Schmitt}
\author{Stefan Kehrein}
\affiliation{%
 Institute for Theoretical Physics, 
	Georg-August-Universit\"at G\"ottingen, 
	Friedrich-Hund-Platz 1, 37077 G\"ottingen, Germany
}
\date{\today}

\maketitle

\section{Extraction of echo peak decay from numerical data}
The analysis of the echo peak decay as shown in Fig.\ 3 of the main text requires particular caution.
In the following, we describe the details of our analysis. 

Each point for $E_O(\tau)$ shown in Fig.\ 3 of the main text is obtained by conducting the following two steps:
\begin{enumerate}
\item Compute the full time evolution with the echo protocol for the given waiting time $\tau$. Since we use the Lanczos algorithm, which results in an iterative propagation of the wave function, we have to compute the full dynamics although we are only interested in the behavior at the echo time. The result is a curve showing an echo peak as depicted in Fig.\ 2 of the main text. Note that for a data point $E_O(\tau)$ time evolution up to $t=2\tau$ is required.
\item From our previous work, Ref.\ [26] of the main text, we know that depending on the choice of the echo protocol the echo peak does not necessarily occur at $t_E=2\tau$. Hence, we extract the echo peak height from the data as denoted in Eq.\ (7) of the main text: we search for the maximal deviation from the saturation value during the backwards propagation. The result gives the echo time $t_E$ and the corresponding echo peak height $E_O(\tau)$.
\end{enumerate}

In the numerical simulation of finite systems we remove the parallel component of the perturbed state, which will vanish in the thermodynamic limit, by hand, meaning that in practice the perturbation operator $\hat P_{\delta t}$ introduced in the main text involves a $\tau$-dependent projection:
\begin{align}
	\hat P_{\delta t}^\tau=\frac{\big(1-\ket{\psi(\tau)}\bra{\psi(\tau)}\big)e^{-\im\hat H_p\delta t}}{\sqrt{1-|\braket{\psi(\tau)|e^{-\im\hat H_p\delta t}|\psi(\tau)}|^2}}
\end{align}
Here $\ket{\psi(\tau)}=e^{-\im\hat H\tau}\ket{\psi_0}$ is the forward-evolved state.

Although the projection onto the orthogonal component of the perturbed state is included in the echo dynamics, the computed echo peak height does not necessarily decay to zero, because at finite size there are additional time-independent contributions in Eq.\ (6) of the main text leading to a non-vanishing long time limit. For the extraction of the decay law of the time-dependent part it is essential to subtract all of the time-independent contributions. In order to estimate the stationary value we compute an average over peak heights at late times. In practice, we computed echos up to time $J\tau=80$ for $\hat H_\text{loc}$ and $J\tau=60$ for $\hat H_\text{fc}$ and averaged over $J\tau\in[60,80]$ or $J\tau\in[40,60]$, respectively, to obtain an estimate of the stationary value that was then subtracted from the result. 

\section{Finite size analysis}
To demonstrate the existence of persisting echos in finite systems and their vanishing in the thermodynamic limit we consider the difference of the echo at $t=2\tau$ from the perfect echo,
\begin{align}
	\Delta M(\tau)=\braket{\psi_0|\hat M|\psi_0}
	-\braket{\psi_0|\hat U_E^\delta t(\tau)^\dagger\hat M\hat U_E^\delta t(\tau)|\psi_0}\ ,
\end{align}
where $\hat U_E^\delta t(\tau)=e^{\im\hat H_p\tau}\hat P_{\delta t}e^{-\im\hat H_p\tau}$. In this case do not project out the parallel component after the perturbation.

\begin{figure}[h!]
\includegraphics{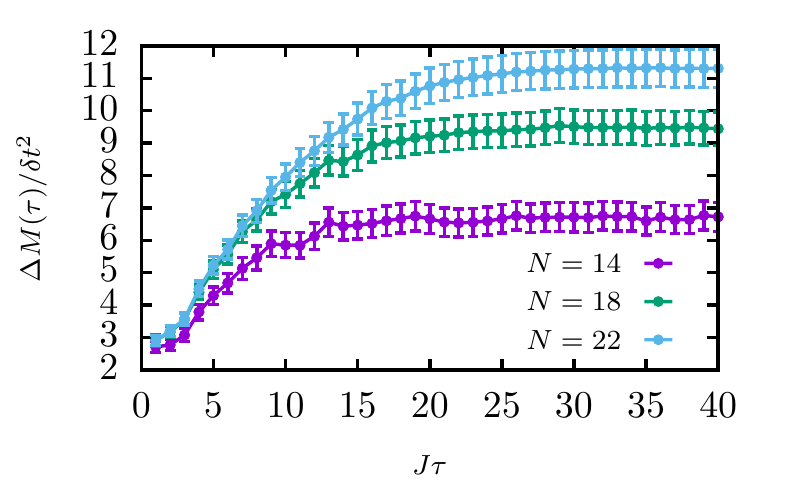}
\caption{Divergence from the perfect echo for different system sizes and $\delta t=0.01/J$.}
\label{fig:finite_size}
\end{figure}

Fig. \ref{fig:finite_size} shows $\Delta M(\tau)$ obtained for $\hat H_\text{loc}$ and averaged over 40 realizations of the perturbation Hamiltonian $\hat H_p$. The saturation of the difference to the perfect echo at late times corresponds to the persistent echo. Clearly, however, the saturation value increases as the system size is increased, indicating that the persistent echo vanishes in the thermodynamic limit $N=\infty$.
